\documentclass[11pt,tightenlines,eqsecnum,floats,aps,amsmath,amssymb,nofootinbib,prd,shownopacs,floatfix]{revtex4-2}

\usepackage{graphicx}
\usepackage{epstopdf}
\usepackage{latexsym}
\usepackage{amssymb}
\usepackage{amsmath}
\usepackage{color}
\usepackage{mathrsfs}
\usepackage{xparse}
\usepackage{float}
\usepackage{mathtools}
\usepackage{multirow}

\usepackage[center]{subfigure}

\begin{document}

  \renewcommand\arraystretch{2}
 \newcommand{\bq}{\begin{equation}}
 \newcommand{\eq}{\end{equation}}
 \newcommand{\bqn}{\begin{eqnarray}}
 \newcommand{\eqn}{\end{eqnarray}}
 \newcommand{\nb}{\nonumber}
 \newcommand{\cb}{\color{blue}}
    \newcommand{\cc}{\color{cyan}}
     \newcommand{\lb}{\label}
        \newcommand{\cm}{\color{magenta}}
\newcommand{\rc}{\rho^{\scriptscriptstyle{\mathrm{I}}}_c}
\newcommand{\rd}{\rho^{\scriptscriptstyle{\mathrm{II}}}_c}
\NewDocumentCommand{\evalat}{sO{\big}mm}{%
  \IfBooleanTF{#1}
   {\mleft. #3 \mright|_{#4}}
   {#3#2|_{#4}}%
}
\newcommand{\PRL}{Phys. Rev. Lett.}
\newcommand{\PL}{Phys. Lett.}
\newcommand{\PR}{Phys. Rev.}
\newcommand{\CQG}{Class. Quantum Grav.}
\newcommand{\parallelsum}{\mathbin{\!/\mkern-5mu/\!}}
\title{Does loop quantum $\mu_o$ scheme permit a black hole formation?}
\author{Bao-Fei Li $^{1}$}
\email{baofeili1@lsu.edu}
\author{Parampreet Singh$^{1}$}
\email{psingh@lsu.edu}
\affiliation{
$^{1}$ Department of Physics and Astronomy, Louisiana State University, Baton Rouge, LA 70803, USA}

\begin{abstract}
We explore the way different loop quantization prescriptions affect the formation of trapped surfaces in the gravitational collapse of a homogeneous dust cloud, with a particular emphasis on the so called $\mu_o$ scheme in which loop quantum cosmology was initially formulated. Its undesirable features  in cosmological models led to the so-called improved dynamics or the $\bar \mu$ scheme. While the jury is still out on the right scheme for black hole spacetimes, we show that as far as the black hole formation is concerned the $\mu_o$ scheme has another, so far unknown,  serious problem.  We find that in the $\mu_o$ scheme no trapped surfaces would form for a non-singular collapse of a homogeneous dust cloud  in the marginally bound case unless the minimum non-zero area of the loops over which holonomies are computed or the Barbero-Immirzi parameter decreases almost four times from its standard value. It turns out that the trapped surfaces in the $\mu_o$ scheme for the marginally bound case are also forbidden for an arbitrary matter content as long as the collapsing interior is isometric to a spatially flat Friedmann-Lema\^itre-Robertson-Walker (FLRW) spacetime.  We find that in contrast to the situation in the $\mu_o$ scheme, black holes can form in the $\bar \mu$ scheme, and also other lattice refinements with a mass gap determined by quantum geometry.  

\end{abstract}
\maketitle

\section{Introduction}

One of the most important predictions of loop quantum cosmology (LQC) is the resolution of the curvature singularity in the Planck regime by replacing it with a quantum bounce \cite{aps,aps2,aps3,acs2010,ashtekar-singh11}. The generic resolution of singularities has been obtained for isotropic as well as anisotropic spacetimes \cite{Singh_2009,Singh:2011gp, Singh:2014fsy, Saini_2017, saini_2018}. The techniques developed in LQC can be readily applied to study the final state of the gravitational collapse too, such as the gravitational collapse of a homogeneous dust cloud whose interior spacetime is described by the  Lema\^itre–Tolman–Bondi  (LTB) metric. An important question in the loop quantization program, as in any other quantization strategy, is to find a preferred quantization which is compatible with physical phenomena and rule out other possible prescriptions. This exercise has been done in past in isotropic  cosmological spacetimes which uniquely selects the so-called $\bar \mu$-scheme or the improved dynamics \cite{cs08,cs09} put forward in \cite{aps3}. This question has also been asked for the static black hole spacetimes \cite{Ashtekar:2018lag, Ashtekar_2018}. But so far there is no comparative exploration on this issue for gravitational collapse using different quantization prescriptions.\footnote{Recently, a question on similar lines was asked for a homogeneous collapse to understand the role of triad vs gauge-covariant fluxes, but only for a specific quantization prescription (the $\bar \mu$ scheme) \cite{Giesel-Li-Singh2021}.}  Thanks to the isometry with the FLRW spacetime for the homogeneous collapse, the physical implications of the different quantization prescriptions mathematically permitted in LQC can be examined in the context of gravitational collapse. Though the results known in cosmological spacetimes directly apply to such a simple gravitational collapse we show this exercise reveals a so far unknown serious problem which is absent in cosmological setting with one of the main loop quantization prescriptions.

Let us recall briefly two most well known loop quantization prescriptions: the $\mu_o$ scheme \cite{abl} and the $\bar \mu$ scheme \cite{aps3}. While the $\mu_o$ scheme can follow naturally from loop quantum gravity (LQG), so far it has been difficult to obtain the $\bar \mu$ scheme. Moreover, in the cosmological setting, the  $\mu_o$ scheme has several undesirable properties \cite{cs08}. Firstly, in the $\mu_o$ scheme, the maximum energy density at which the bounce occurs is found to be dependent on the phase space variables whose initial conditions can be chosen such that the bounce can happen at classical densities. Secondly, in the case of non-compact manifolds the dynamics depends on the choice of the fiducial cell used to define the symplectic structure. Finally, in presence of matter which violates the strong energy condition the universe recollapses at small spacetime curvatures. All of these undesirable features arise from the way $\mu_o$ scheme is defined where $\mu_o$ measures edge length of the holonomies and is put equal to the square root of the minimum area eigenvalue in loop quantum gravity. These problems are absent in the $\bar \mu$ scheme where one considers the physical area of the loop to determine the value of edge length $\bar \mu$. In the context of the gravitational collapse of a dust cloud, the interior spatial manifold is compact and strong energy condition is not violated. Hence, two of the problems of using $\mu_o$ scheme are eliminated. Though the problem of bounce density depending on phase space variables remains, one can engineer a small set of initial data such that the bounce density is in the Planck regime.

The strategy used in defining $\mu_o$ and $\bar \mu$ schemes in LQC has led to introduction of various quantization prescriptions for loop quantization of static (vacuum) black hole spacetimes which can be viewed as generalizations of above schemes (see for eg.
 \cite{Ashtekar_2005, Bohmer_2007, Campiglia_2008,Gambini_2008,Gambini_2013,Gambini2014,Corichi_2016, Olmedo:2017lvt,Ashtekar:2018lag, Ashtekar_2018,Gambini_2020}).
  Investigations on singularity resolution have been carried out also for   dynamical gravitational collapse spacetimes when a matter field is taken into account\footnote{For early works taking only inverse volume modifications into account, see  \cite{Bojowald-singh2005,Goswami_2006}. These modifications only become relevant near Planck lengths and are negligible compared to holonomy modifications \cite{aps2,aps3}. In this work, we ignore the inverse volume modifications.} \cite{Gambini_2009,Bambi_2013,Tavakoli_2014,Ben_tez_2020,han2020improved, Kelly_2,Giesel-Li-Singh2021}. These studies mainly focused on the $\bar \mu$ scheme  and showed that the central singularity is generically resolved and a black hole can form when its mass is above some threshold value.  In this manuscript, we study the $\mu_o$ scheme in the simplest setting where a homogeneous dust cloud is collapsing in the marginally bound case. In order to highlight the main features of the $\mu_o$ scheme in the homogeneous gravitational collapse, we also consider other loop quantizations due to different kinds of lattice refinement, including the $\bar \mu$ scheme and others parameterized in \cite{cs08}. We find that although central singularity is generically resolved and replaced with a bounce in all loop quantization schemes considered in this paper, regardless of the initial conditions and even the matter content, the $\mu_o$ scheme can not allow for the formation of the trapped surface as long as the minimal eigenvalue of the area operator which directly determines the specific value of $\mu_o$ is given by LQG. In contrast, for other loop quantizations, including the $\bar \mu$ scheme, the trapped surface can always form during the collapse of the dust cloud when the dust mass is larger than a threshold mass whose order of magnitude is dependent on the lattice refinement.
  We show that the trapped surface can not form in the $\mu_o$ scheme even for arbitrary matter which is compatible with the isometry with FLRW interior. For trapped surface to form in the $\mu_o$ scheme the minimum non-zero eigenvalue of the area operator or the Barbero-Immirzi parameter must decrease almost four times. Further, such a value of minimum non-zero area eigenvalue or the Barbero-Immirzi parameter would violate the covariant entropy bound.

In the following, we first briefly review of the homogeneous dust shell model in Sec. \ref{sec:review}, including the classical Hamiltonian constraint and the matching conditions between the interior spacetime and the exterior generalized Vaidya spacetime. In Sec. \ref{sec:quantizations}, we first discuss the loop quantization of the interior collapsing spacetime in the $\mu_o$ scheme and its physical consequences. Then we proceed with other loop quantizations including the $\bar \mu$ scheme and other possible lattice refinements. In this section, we concentrate on the role of different loop quantizations on the formation of the trapped surface during the non-singular evolution of the dust cloud. Finally, we conclude our main results in Sec. \ref{sec:outlook}. Throughout this paper, we use the Planck units in which $\hbar=c=1$ while keeping Newton's constant $G$ explicit in our formulae.

\section{Preliminaries: the classical dust shell model}
\lb{sec:review}

In this section, we briefly review the classical dust shell model of the gravitational collapse in the marginally bound case using connection and triad variables. Our discussion would parallel the one in \cite{Giesel-Li-Singh2021}, which we refer the reader for further details.

The dust shell model applies to a homogeneous evolution of the dust cloud which ignores the interactions between neighboring shells and thus leads to a collapse of all of the dust shells  at a uniform speed. As a result, the dynamics of the whole dust cloud can be inferred from the dynamics of the outermost dust shell. The interior collapsing spacetime is described by the classical LTB metric,
\bq
\label{metricLTB}
ds^2_- = -dt^2 +\frac{(R^\prime)^2}{1+2f(x)}dx^2+R^2d\Omega^2,
\eq
where $x$ is the radial coordinate, $R$ is the areal radius and $d\Omega^2=d\theta^2+\sin^2\theta d\phi^2$ is the angular part of the metric. The prime denotes the differentiation with respect to $x$. The function $f$ only depends on the radial coordinate $x$ and stands for  the total energy per unit mass at $x= const$. Depending on the sign of $f$, there exist three distinct cases: the marginally bound case with $f=0$, the bound case with $f<0$ and the unbound case with $f>0$. In all three cases, the classical dynamical evolution is governed by
\bq
\lb{dynamical equations}
8\pi G\rho_{\rm dust} =  \frac{F^\prime}{R^2R^\prime}\quad{\rm and}\quad
\dot{R}^2=\frac{F}{R}+2f,
\eq
where an overdot denotes differentiation with respect to the proper time $t$, $\rho_{\rm dust}$ denotes the energy density of the dust cloud and $F/2G$ which only depends on the radial coordinate $x$ is the active gravitating mass enclosed within the dust sphere with radius $R(x)$.

In the following, we only focus on the homogeneous collapse of the dust cloud in the marginally bound case with $f=0$. The homogeneous evolution of the dust cloud implies that the energy density of the dust cloud only depends on the proper time and consequently the areal radius takes the form $R=xa(t)$. As a result, the interior LTB metric (\ref{metricLTB}) reduces to the one for a spatially flat FLRW universe.
Correspondingly, the dynamical equations in (\ref{dynamical equations}) are equivalent to the equation
\bq
\lb{classical Friedmann flat case}
\left(\frac{\dot R }{R}\right)^2=\frac{8\pi G}{3}\rho_\mathrm{dust}.
\eq
As already discussed in detail in \cite{Giesel-Li-Singh2021}, in terms of the areal radius $R$ and its conjugate momentum,  the gravitational and the matter sectors of the Hamiltonian constraint of the outermost dust shell are given respectively by
\bq
\label{shellHam}
 H^\mathrm{shell}_{\rm grav} = -\frac{\dot{R}_b^2R_b}{2G}\quad{\rm and}\quad  H^\mathrm{shell}_{\rm dust} = \frac{F_b}{2G},
\eq
where the angular part in the Hamiltonian constraint has already been integrated and the integration along the radial direction has been performed  from the center of the dust cloud $x=0$ to its outermost shell $x=x_b$ (namely the boundary surface), yielding $R_b=a(t)x_b$ and $F_b=F(x_b)$. In order to loop quantize the dust shell model, it is a prerequisite to express the above Hamiltonian constraint of the dust shell model in terms of the Ashtekar variables which was first analyzed in \cite{Bojowald:2008ja} for an inhomogeneous dust cloud and later adapted to the shell model for a homogeneous dust cloud in \cite{Giesel-Li-Singh2021}. In the dust shell model, one can identify the radial component of the densitized triad with the square of the areal radius, namely $R_b\coloneqq\sqrt{|E^x_b|}$
with $E^x_b\coloneqq E^x(x_b)$. Then, in terms of $E^x_b$ and its conjugate momentum $K_{x_b}\coloneqq K_x(x_b)$ which is half of the radial component of the extrinsic curvature, the Hamiltonian of the outermost shell of the dust cloud can be written as \cite{Giesel-Li-Singh2021}
\bq
\lb{classical-shell-ham}
H^\mathrm{shell}_\mathrm{classical}=-\frac{2}{G} K_{x_b}^2\sqrt{|E^x_b|}=-\mathcal E_\mathrm{dust},
\eq
with $F_b=2G\mathcal E_\mathrm{dust}$ and $\mathcal E_\mathrm{dust}$ represents the mass of the dust cloud. Finally, one can recover the classical Hamiltonian constraint in the cosmological setting for the dust interior by identifying
\bq
K_{x_b}=\frac{c}{2\gamma}\left(\frac{3}{4\pi}\right)^{1/3},\quad \quad E^x_b=p\left(\frac{3}{4\pi}\right)^{2/3},
\eq
in which eq. (\ref{classical-shell-ham}) reduces to the familiar form in terms of the connection $c$ and the triad $p$, namely
\bq
\lb{classical Hamiltonian}
H^\mathrm{shell}_\mathrm{classical}=-\frac{3\sqrt{p} c^2}{8\pi G\gamma^2}+\mathcal E_\mathrm{dust},
\eq
with $\{c,p\}=\frac{8\pi G\gamma}{3}$ and the volume of the dust interior is given by $V=p^{3/2}=\frac{4\pi}{3}R^3_b$.

For a complete description of the gravitational collapse of the dust cloud, it is necessary to match the interior collapsing spacetime with an exterior spacetime. Without loss of generality, we choose the generalized Vaidya spacetime as the exterior spacetime since in general the dust cloud acquires an effective non-vanishing pressure after quantum gravity effects are taken into account. The generalized Vaidya spacetime in the advanced Eddington-Finkelstein coordinates takes the form \cite{Israel1991}
\bq
\label{Vaidya}
ds^2_+=-\left(1-\frac{2GM(\nu,Y)}{Y}\right)d\nu^2+2d\nu dY+Y^2d\Omega^2,
\eq
where $M(\nu,Y)$ stands for the Vaidya mass. Note the interior spacetime (\ref{metricLTB}) is glued with the exterior (\ref{Vaidya}) by requiring the continuation of the first and second fundamental forms at the boundary surface $\Sigma_b: x=x_b$.
The matching conditions can be worked out explicitly by computing the first and second fundamental forms at $\Sigma_b$ from the interior and the exterior metrics, yielding \cite{goswami-joshi04}
\bqn
\label{matching conditions}
\evalat[\big]{Y}{\Sigma_b}&=&R_b(t)=x_b a(t),\quad \quad \evalat[\bigg]{\left(\frac{d\nu}{dt}\right)}{\Sigma_b}=\frac{R_{,x}+x_b\dot a}{1-F/R},\\
\label{matching conditions 2}
F_b(t)&=&2M(\nu,Y)G,\quad  \quad  M(\nu,Y)_{,Y}G=\frac{F}{2R}+x^2_ba\ddot a .
\eqn
As a result, the boundary surface would become a trapped surface when $2GM(\nu,Y)>R_b$ or equivalently $F_b>R_b$. Using the second dynamical equation in (\ref{dynamical equations}), the criterion for the formation of the trapped surface in the marginally bound case turns out to be $\dot R^2_b>1$. Moreover, from the detailed discussions on the expansion parameters given in \cite{Giesel-Li-Singh2021}, we can identify a black hole for $\dot R_b<-1$ in the contracting phase  and a white hole for $\dot R_b>1$ in the expanding phase. Classically, these two branches are disjoint, and one inevitably encounters a central singularity either in the past or in the future. However, after loop quantization, the expanding and contracting branches can be  connected via a bounce due to quantum gravity effects, such as in the $\bar \mu $ scheme \cite{Giesel-Li-Singh2021}.

\section{Loop quantizations of the dust shell model}
\lb{sec:quantizations}

In this section, we discuss the physical implications of the  loop quantizations of the dust shell model in the marginally bound case with an emphasis on the question of formation of the trapped surfaces in the $\mu_o$ scheme.  Our analysis is based on the effective dynamics which is proved via numerical simulations to faithfully capture the underlying quantum dynamics for the states sharply peaked around the classical solutions at  late times \cite{aps3,numlsu-1,numlsu-2,numlsu-3}.

\subsection{The $\mu_o$ scheme}
The loop quantization of the classical Hamiltonian constraint  is carried out in terms of the Ashtekar-Barbero connection and the densitized triad which in the homogeneous spacetime reduce to the connection $c$ and the triad $p$. The non-perturbative quantum gravity modification arises from the regularization of the field strength of the connection which leads to a quantum difference equation whose dynamics can be faithfully captured by an effective Hamiltonian constraint for the sharply peaked states. For the classical Hamiltonian constraint of the dust shell model given in (\ref{classical Hamiltonian}),  the effective Hamiltonian constraint in the $\mu_o$ scheme takes the form \cite{aps2}
\bq
\lb{effective Hamiltonian in spatially flat universe}
\mathcal H^\mathrm{(\mu_o)}_\mathrm{eff}=-\frac{3\sqrt p}{8\pi G \gamma^2\mu^2_o}\sin^2\left(\mu_o c\right)+\mathcal E_\mathrm{dust} .
\eq
Using the Hamilton's equation for the triad $p$ and
the vanishing of the effective Hamiltonian constraint one obtains
\bq
\lb{dynamics1}
\left(\frac{\dot R_b}{R_b}\right)^2=\frac{8\pi G}{3}\rho_\mathrm{dust}\left(1-\frac{8\pi G \gamma^2\mu^2_op}{3}\rho_\mathrm{dust}\right).
\eq
The quadratic term with a negative sign tells us that $\dot R_b/R_b$ vanishes at a maximum energy density
%
\bq
\rho^\mathrm{(\mu_o)}_\mathrm{max}=\frac{27}{(8\pi G\gamma^2\mu^2_o)^3\mathcal E^2_\mathrm{dust}}.
\eq
Unlike the classical theory, the central singularity is avoided in the collapse of the dust shell. Instead the shell bounces at $\rho = \rho^\mathrm{(\mu_o)}_\mathrm{max}$.
Note the maximum energy density in the $\mu_o$ scheme is not a constant but depends explicitly on the mass of the dust cloud. When the dust mass is taken to be macroscopic, the bounce takes place at very small spacetime curvature. This problem is well known in the cosmological setting for the $\mu_o$ scheme \cite{aps2,cs08}. Our goal here is to investigate the formation of trapped surfaces. For this it
 is sufficient to  note that the square of the velocity of the outermost shell is given by
\bq
\lb{velocity squared in k=0}
\dot {R}^2_b=\left(\frac{3}{4\pi}\right)^{2/3}\frac{\dot p^2}{4p}=\left(\frac{3}{4\pi}\right)^{2/3}x\left(1-\gamma^2 \mu^2_o x\right),
\eq
where $x$ is defined by
\bq
x\coloneqq \frac{8\pi G\mathcal E_\mathrm{dust}}{3\sqrt{p}} ~.
\eq
On the other hand, from the effective Hamiltonian constraint (\ref{effective Hamiltonian in spatially flat universe}), one can find
\bq
\lb{a1}
x=\frac{\sin^2\left(\mu_o c\right)}{ \gamma^2\mu^2_o}\in\left(0,\frac{1}{ \gamma^2\mu^2_o}\right).
\eq
It is important to note that the above upper and lower limits for $x$ are constant which do not depend on the triad $p$. As a result, combining (\ref{velocity squared in k=0}) and (\ref{a1}), we find at $x=\frac{1}{2\gamma^2 \mu^2_o}$, the square of the velocity of the outermost dust shell attains its maximum value which turns out to be
\bq
\lb{max velocity munot}
{\dot R^2}_\mathrm{max}=\left(\frac{3}{4\pi}\right)^{2/3}\times\frac{1}{4\gamma^2\mu^2_o}\approx0.063<1,
\eq
where we have used $\mu_o=3\sqrt 3\approx5.20$ and $\gamma\approx0.2375$ which is fixed by the black hole thermodynamics discussed in \cite{Meissner_2004}. For these values of $\mu_o$ and $\gamma$ a trapped surface can not form in the $\mu_o$-scheme. Can one choose other values of $\gamma$ and $\mu_o$ such that $\dot R_b^2 > 1$?
In order to have a black hole form during the collapse, the factor $\gamma\mu_o$ in the denominator of (\ref{max velocity munot}) has to satisfy $\gamma \mu_o\lesssim 0.31$ which implies we need to decrease either  $\gamma$ and/or $\mu_o$. The freedom to choose a different $\mu_o$ depends on how $\mu_o$ is determined in the $\mu_o$ scheme. As discussed in \cite{aps2}, $\mu_o$ is related with the minimal non-zero eigenvalue of the area  operator  denoted by $\Delta$ via
\bq
\mu_o=\frac{6\Delta}{8\pi \gamma \ell^2_\mathrm{pl}},
\eq
where $\ell_\mathrm{pl}$ is the Planck length. Since $\Delta$ is proportional to $\gamma$, decreasing the  Barbero–Immirzi parameter does not affect the the magnitude of $\mu_o$. If the value of $\mu_o$ needs to be decreased then one would have to change the way minimum area of the loop over which holonomies are computed is equated with the minimum non-zero area eigenvalues in LQG. If one keeps Barbero–Immirzi parameter unchanged, the value of $\mu_o$ has to decrease by four times. On the other hand, if $\mu_o$ is kept fixed, one needs to decrease the Barbero-Immirzi parameter by four times. But, the value of the Barbero-Immirzi parameter is determined by the black hole entropy. Though the value $\gamma \approx 0.2375$ is generally used in LQC \cite{Meissner_2004}, other evaluations exist
 \cite{Ashtekar_1998,Dreyer_2003,Domagala_2004,Ansari_2008} and the lowest evaluated value is $\gamma \approx \frac{\ln3}{\sqrt 8\pi}\approx0.12$ \cite{Dreyer_2003}. Even if one uses this value, the $\mu_o$ scheme does not allow a trapped surface formation. Interestingly, in earlier works in LQC,  $\Delta$ was chosen to be the lowest non-zero eigenvalue $2\sqrt 3 \pi \gamma \ell^2_\mathrm{pl}$ but its corresponding eigenstates were argued to not yield homogeneous classical metrics \cite{Ashtekar_2009}. Despite a lack of physical justification to examine this, even using this value of $\mu_o$ along with $\gamma \approx \frac{\ln3}{\sqrt 8\pi}\approx0.12$ does not allow formation of trapped surfaces since
one finds
$\gamma \mu_o\approx0.32$ which is still larger than the value ($\approx0.31$). Further, such a change in the value of the Barbero-Immirzi parameter is incompatible with the covariant entropy bound. Here we note that the value $\gamma \approx 0.2375$ which is currently used in the literature leads to a maximum energy density in the $\bar \mu$ scheme which almost saturates the covariant entropy bound \cite{Ashtekar_2008}. Any decrease in $\gamma$, as needed above, would not be consistent with the covariant entropy bound in LQC. Since $\gamma$ should be a universal constant in loop quantizations, irrespective of whether one chooses $\mu_o$ or $\bar \mu$ scheme, we conclude that if trapped surfaces form in the $\mu_o$ scheme for the model discussed above, the covariant entropy bound would be violated. In conclusion, irrespective of the initial conditions, the black hole would not form during the collapse of a homogeneous dust cloud in the marginally bound case for any possibly allowed values of $\mu_o$ and $\gamma$ which have appeared in the literature. \\

\noindent
{\bf Remarks}: It should be noted that the result that no trapped surface can form in the marginally bound case with the $\mu_o$ scheme can be extended to an arbitrary matter content. The effective Hamiltonian constraint for any matter content can be expressed as
\bq
\mathcal H^\mathrm{(\mu_o)}_\mathrm{eff}=-\frac{3\sqrt p}{8\pi G \gamma^2\mu^2_o}\sin^2\left(\mu_o c\right)+\mathcal H_m,
\eq
where $\mathcal H_m$ stands for the matter Hamiltonian. It is straightforward to check from the corresponding Hamilton's equations, that one obtains the same expression for $\dot R^2$ as given in (\ref{velocity squared in k=0}) with $x\coloneqq \frac{8\pi G\mathcal H_m}{3\sqrt{p}}$ which takes the values in the same range shown in (\ref{a1}). As a result, for any matter content, the maximum velocity squared of the outermost shell of a collapsing cloud is still given by (\ref{max velocity munot}) which implies no trapped surface can form during the collapse and the expansion of the cloud, irrespective of the initial conditions and the matter content.

\subsection {The $\bar \mu$ scheme}
The effective dynamics in the $\bar \mu$ scheme is governed by the effective Hamiltonian constraint, which takes the form \cite{aps3}
\bq
\lb{effective Hamiltonian in mubar}
\mathcal H^\mathrm{(\bar \mu)}_\mathrm{eff}=-\frac{3 p^{3/2}}{8\pi G \gamma^2\lambda^2}\sin^2\left(\lambda \frac{c}{\sqrt p}\right)+\mathcal E_\mathrm{dust},
\eq
where $\lambda =\sqrt \Delta$ and the resulting dynamics has already been discussed in detail in the context of the gravitational collapse in \cite{Giesel-Li-Singh2021}. The corresponding equation for the outermost shell takes the form
\bq
\lb{Friedmann mubar}
\left(\frac{\dot R_b}{R_b}\right)^2=\frac{8\pi G}{3}\rho_\mathrm{dust}\left(1-\frac{\rho_\mathrm{dust}}{\rho_c}\right),
\eq
where $\rho_c=\frac{3}{8\pi G\gamma^2\lambda^2}\approx 0.41 m^4_\mathrm{pl}$. As a result, the bounce in the $\bar \mu$ scheme takes place in the Planck regime at a fixed maximum energy density which does not depend on the initial conditions. Furthermore, during the evolution of the dust cloud, the velocity squared of the outermost dust shell is given by
\bq
\lb{velocity-squared}
\dot R^2_b=\frac{8\pi G \mathcal E_\mathrm{dust}^{2/3}\rho^{1/3}_\mathrm{dust}}{(48\pi^2)^{1/3}}\left(1-\frac{\rho_\mathrm{dust}}{\rho_c}\right).
\eq
Therefore, at $\rho_\mathrm{dust}=\rho_c/4$, $\dot R^2_b$ attains its maximum value which turns out to be
\bq
\label{3a4}
\dot R^2_\mathrm{max}=\frac{3}{4}\left(\frac{G\mathcal E_\mathrm{dust}}{\lambda\gamma }\right)^{2/3}.
\eq
The above equation implies that in contrast  to the $\mu_o$ scheme, the dust mass would affect the formation of  the trapped surface in the $\bar \mu$ scheme. To be specific, we find that there exists a threshold dust mass \cite{Giesel-Li-Singh2021}
\bq
\mathcal E^*_\mathrm{dust}=\frac{8\lambda\gamma}{3\sqrt 3 G}\approx0.831 m_\mathrm{pl} ~.
\eq
When the initial dust mass is larger than $\mathcal E^*_\mathrm{dust}$, the trapped surface would form during the collapse or the expansion of the dust cloud. On the other hand, if the initial dust mass is lower than $\mathcal E^*_\mathrm{dust}$, no trapped surface can form during the non-singular evolution of the dust cloud.

\subsection{ Other quantizations with different kinds of lattice refinement}

\begin{figure}
{
\includegraphics[width=8cm]{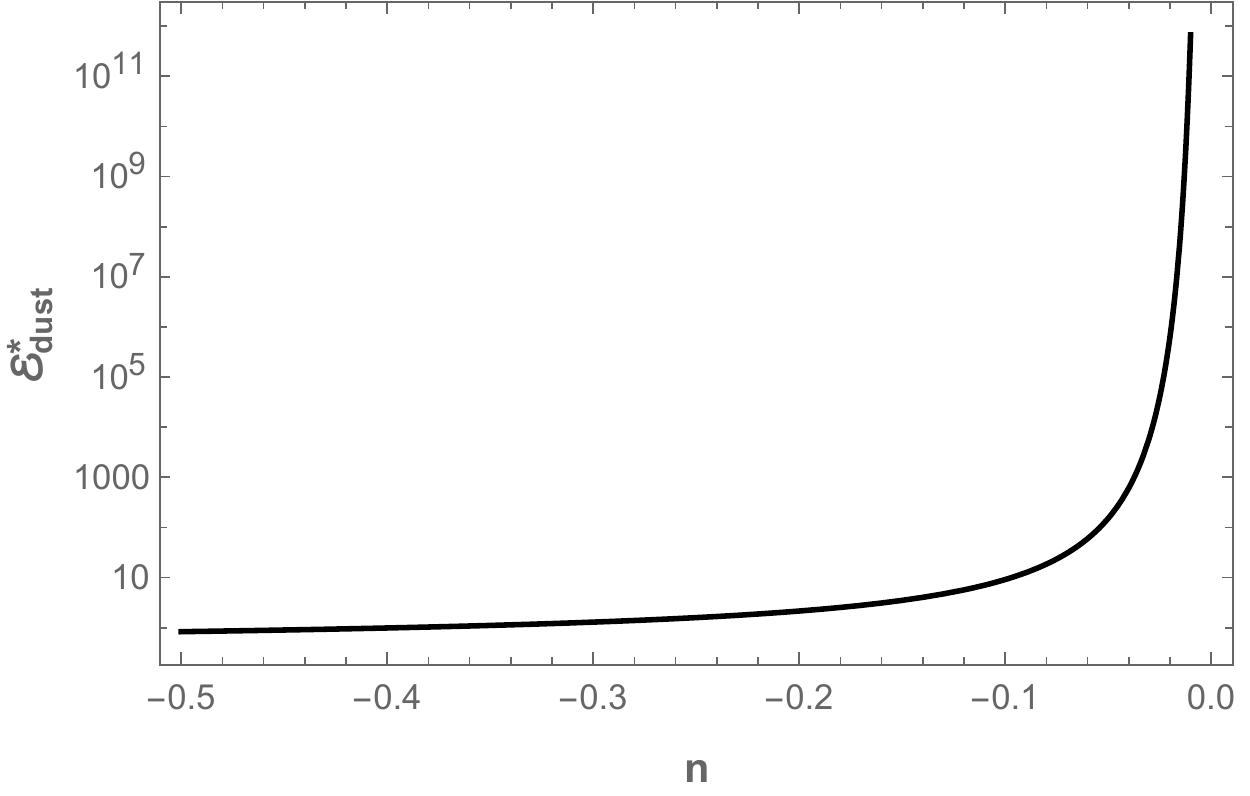}
}
\caption{In the figure, we show the qualitative relationship between the threshold mass and the parameter $n$ by assuming that $\mu_n$ is numerically of a similar order as $\mu_o$ ($n=0$) or $\lambda$ ($n=-1/2$). To be specific, in the plot we choose the numerical value of $\mu_n$ same as that of $\lambda$. The threshold mass for the formation of the trapped surface tends to be close to the Planck mass as $n$ approaches $-1/2$ and it approaches infinity when $n$ approaches zero.  }
\label{fig1}
\end{figure}

In literature it has been sometimes argued that one can have quantization prescriptions based on  lattice refinements which are a more general case of $\mu_o$ and $\bar \mu$ schemes \cite{Bojowald2006}. Though the physical derivation of such refinements remains unclear, their implications have been discussed earlier in the cosmological setting which rules all of them out except the $\bar \mu$ scheme \cite{cs08}. We consider these generalizations for the formation of trapped surfaces. In this model, one deals with a generalized set of the phase space variables which are obtained from $(c,p)$ variables by a canonical transformation \cite{cs08}
\bq
P_g=cp^n,~~~g=\frac{p^{1-n}}{1-n},
\eq
where $n$ is a constant which carries the information of a particular refinement and can take any values in the range $[-1/2,0]$. The case $n= 0$ corresponds to the $\mu_o$ scheme and $n= -1/2$ corresponds to the $\bar \mu$ scheme.  Following the ideas in loop quantization, one then polymerizes the momentum $P_g$, leading to the effective Hamiltonian constraint \cite{cs08}
\bq
\mathcal H_\mathrm{eff}=-\frac{3p^{\frac{1-4n}{2}}}{8\pi G \gamma^2 \mu^2_n}\sin^2\left(\mu_n cp^n\right)+\mathcal E_\mathrm{dust},
\eq
where $\mu_n$ only depends on $n$.  From the effective Hamiltonian constraint, one can derive the corresponding Hamilton's equations and a constraint equation for $R_b$ which takes the same form as (\ref{Friedmann mubar}) but with a different maximum energy density given explicitly by
\bq
\rho_\mathrm{max}=\frac{3}{8\pi G\gamma^2\mu^2_n}\left(\frac{8\pi G\gamma^2\mu^2_n\mathcal E_\mathrm{dust}}{3}\right)^{\frac{4n+2}{4n-1}}.
\eq
As a result, only in the $\bar \mu$ scheme, the maximum energy density does not depend on the dust mass which agrees with the observations in \cite{cs08}. For other lattice refinements including the $\mu_o$ scheme, the maximum energy density is always determined by the dust mass. On the other hand, it is straightforward to obtain the velocity squared of the outermost shell of the dust cloud for an arbitrary $n$, which takes the form
\bq
\dot R^2_b=\left(\frac{3}{4\pi}\right)^{2/3}\times\frac{8\pi G\mathcal E_\mathrm{dust}}{3\sqrt p}\left(1-\frac{8\pi G\gamma^2\mu^2_n\mathcal E_\mathrm{dust}}{3p^{\frac{1-4n}{2}}}\right).
\eq
Using the Hamiltonian constraint and the definition of the dust energy density, we find the maximum of the velocity squared in terms of the dust mass and the lattice refinement factor $n$,
\bq
\lb{max velocity}
\dot R^2_\mathrm{max}=\left(\frac{3}{4\pi}\right)^{2/3}\left(\frac{8\pi G}{3}\mathcal E_\mathrm{dust}\right)^{-\frac{4n}{1-4n}}\left(\frac{1-4n}{2-4n}\right)\Bigg\{\frac{1}{(2-4n)\gamma^2\mu^2_n}\Bigg\}^{\frac{1}{1-4n}},
\eq
which reduces to eq. (\ref{max velocity munot}) in the $\mu_o$ scheme and eq. (\ref{3a4})  in the $\bar \mu$ scheme. In order to form a trapped surface during the evolution of the dust cloud, one requires the maximum value of the velocity squared larger than unity, which gives rise to the condition
\bq
\lb{threshold mass}
\mathcal E_\mathrm{dust}\ge \mathcal E^*_\mathrm{dust} =\frac{3}{8\pi G}\left(\frac{4\pi}{3}\right)^{-\frac{1-4n}{6n}}\left(\frac{2-4n}{1-4n}\right)^{-\frac{1-4n}{4n}}\Bigg\{\left(2-4n\right)\gamma^2\mu^2_n\Bigg\}^{-\frac{1}{4n}}.
\eq
Except for $n=0$ (the $\mu_o$ scheme), the right hand side of the above equation gives the threshold dust mass $\mathcal E^*_\mathrm{dust}$ for an arbitrary choice of $n$. We show the dependence of the threshold mass on the parameter $n$ in Fig. \ref{fig1}. Assuming that numerically $\mu_n$ takes a value of the same order as $\mu_o$ or $\lambda$, we see from the figure that when $n$ increases from $-1/2$ to zero, the threshold mass increases rapidly and approaches infinity as $n$ approaches zero.  Thus, we find that except in the $\mu_o$ scheme, one can always choose the dust mass so that a trapped surface forms which can be identified with a dynamical black hole or white hole. It is important to note that even though there exist values of $n$ except $n=-1/2$ which allow a formation of the trapped surface, this does not imply their viability. These quantization prescriptions would suffer from dependence of energy density at the bounce on the phase space variables, as a result of which the bounce density can be very small for a gravitational collapse of a macroscopic dust cloud.

\section{Conclusions}
\lb{sec:outlook}

Th main goal of this paper was to explore whether different loop quantization prescriptions allow the formation of a black hole. Our work is based on investigating evolution of a homogeneous dust cloud whose interior is isometric to a spatially flat FLRW spacetime. The homogeneity of the interior spacetime allows us to use the quantization prescriptions used in LQC. In particular, we focused on the $\mu_o$ scheme and compared the results with the $\bar \mu$ scheme and other possible lattice refinements. While the $\mu_o$ scheme already faces three serious challenges in the cosmological setting, two of these, dependence of physical predictions on the choice of a  fiducial cell and recollapse at late times when strong energy condition is violated, are not relevant for the case of a gravitational collapse of a dust cloud.  The third limitation is that bounce can occur at very small energy densities remains, but one can also choose initial conditions of the matter content such that the bounce occurs at Planckian densities. In this manuscript, our objective was not to show that this problem can be resolved for the $\mu_o$ scheme but to add another problem to this list. In summary the problem is that unless the minimum area gap or the Barbero-Immirzi parameter  decreases by almost four times, no trapped surfaces can form in the homogeneous collapse of a dust shell or in fact any other form of matter in the $\mu_o$ scheme. This serious problem was so far unknown and in some sense shows that existence of black holes completely rules out the $\mu_o$ scheme.

In contrast to the situation in $\mu_o$ scheme, we find that trapped surfaces can form in the $\bar \mu$ scheme, a result derived earlier in \cite{Tavakoli_2014,Giesel-Li-Singh2021}, and also for other lattice refinements parameterized by a parameter $n$ following \cite{cs08}. The central singularity for all values of $n$ and all lattice refinements except the case of $n=0$ result in formation of trapped surfaces. Unlike other values of $n \neq 0$, the formation of the trapped surfaces in the $\mu_o$ scheme only depends on the values of $\mu_o$ and the Barbero-Immirzi parameter $\gamma$.  For all of the values of these two parameters that are allowed in the literature, we find no trapped surface can form in the  $\mu_o$  scheme.  As a result, the $\mu_o$ scheme is disfavored in this context.  On the other hand, for other loop quantizations, there always exists some particular threshold mass. When the dust mass is greater than this threshold value, the trapped surfaces can form during the non-singular evolution of the dust cloud. It is remarkable to note that only in the $\mu_o$ scheme, the trapped surface would not form regardless of the dust mass, the initial radius and even the matter content.

A few remarks regarding our main findings are in order.  Firstly,  the exclusion of trapped surfaces in the $\mu_o$ scheme can also be argued using Bousso's covariant entropy bound. If this bound is satisfied the minimal area gap can not  decrease to the value required for the formation of the trapped surface in the $\mu_o$ scheme. Secondly, though we believe that the simplest treatment of the dust collapse in the homogeneous setting does grasp some generic features of the full quantum gravity effects, including the resolution of the central singularity and the lower bounds on the dust mass for the formation of the trapped surfaces in some loop quantizations, since the scale of any finite celestial bodies can not be compared with the universe, a homogeneous reduction which can be well justified for the entire universe, may not be sufficient for describing a realistic gravitational collapse. In other words, to completely exclude the $\mu_o$ scheme, one still needs to further investigate the inhomogeneous case where the energy density also changes in the radial direction and the interactions between neighboring shells could in principle change the dynamics of the outermost shell. Thirdly,  even in the homogeneous case, we only considered the marginally bound case of the dust collapse. Note that the marginally bound case could be matched to the spatially-flat interior and it was straightforward to deal using effective dynamics of spatially-flat isotropic model in LQC. However, an important question is whether these results hold in the presence of spatial curvature, i.e. the bound and the unbound cases. Since the spatial curvature can result in qualitatively distinct dynamics even in the classical theory, its inclusion at the quantum level for the gravitational collapse of a dust cloud  can potentially give rise to richer phenomenology. As an example, the bound case corresponds to the positively curved interior which brings in various non-trivialities under loop quantization because of the interaction between the spatial curvature and quantum geometry effects. It can be shown that for the bound case the $\bar \mu$ scheme yields a trapped surface for different loop quantizations, but the $\mu_o$-scheme fails to give a viable macroscopic model of gravitational collapse, for the holonomy \cite{apsv} and connection based approaches \cite{ck1,ds1}, because of the lack of the classical limit \cite{li-ps}. These are important avenues to be investigated in future for bound and unbound cases and check the robustness of results obtained in this manuscript. Finally, our analysis leaves open the question of whether the generalizations of the $\mu_o$ and $\bar \mu$ schemes which are applied to the static black hole spacetimes  still remain valid for the collapsing spacetime. It will be interesting to apply the arguments used in this manuscript to examine the viability of such approaches.

\section*{Acknowledgments}

This work is supported by the NSF grants PHY-1454832 and PHY-2110207, and the National Natural Science Foundation of China (NNSFC) with the Grants No. 12005186.


\begin{thebibliography}{10}

\bibitem{aps}
A.~Ashtekar, T.~Pawlowski, and P.~Singh,
\newblock {\em Quantum nature of the Big Bang},
\newblock Phys. Rev. Lett. {\bf 96}, 141301 (2006).

\bibitem{aps2}
A.~Ashtekar, T.~Pawlowski, and P.~Singh,
\newblock {\em Quantum nature of the Big Bang: An analytical and numerical
  investigation. I.},
\newblock Phys. Rev. {\bf D73}, 124038 (2006).

\bibitem{aps3}
A.~Ashtekar, T.~Pawlowski, and P.~Singh,
\newblock {\em Quantum nature of the Big Bang: Improved dynamics},
\newblock Phys. Rev. {\bf D74}, 084003 (2006).

\bibitem{acs2010}
A.~Ashtekar, A.~Corichi, and P.~Singh,
\newblock {\em Robustness of key features of loop quantum cosmology},
\newblock Phys. Rev. {\bf D77}, 024046 (2010).

\bibitem{ashtekar-singh11}
A.~Ashtekar and P.~Singh,
\newblock {\em Loop Quantum Cosmology: A Status Report},
\newblock Class. Quant. Grav. {\bf 28}, 213001 (2011).

\bibitem{Singh_2009}
P.~Singh,
\newblock {\em Are loop quantum cosmos never singular?},
\newblock Class. Quant. Grav. {\bf 26}, 125005 (2009).

\bibitem{Singh:2011gp}
P.~Singh,
\newblock {\em Curvature invariants, geodesics and the strength of
  singularities in Bianchi-I loop quantum cosmology},
\newblock Phys. Rev. {\bf D85}, 104011 (2012).

\bibitem{Singh:2014fsy}
P.~Singh,
\newblock {\em Loop quantum cosmology and the fate of cosmological
  singularities}, 2014, arXiv:1509.09182.

\bibitem{Saini_2017}
S.~Saini and P.~Singh,
\newblock {\em Resolution of strong singularities and geodesic completeness in
  loop quantum Bianchi-II spacetimes},
\newblock Class. Quant. Grav. {\bf 34}, 235006 (2017).

\bibitem{saini_2018}
S.~Saini and P.~Singh,
\newblock {\em Generic absence of strong singularities in loop quantum
  Bianchi-IX spacetimes},
\newblock Class. Quant. Grav. {\bf 35}, 065014 (2018).

\bibitem{cs08}
A.~Corichi and P.~Singh,
\newblock {\em Is loop quantization in cosmology unique?},
\newblock Phys. Rev. {\bf D78}, 024034 (2008).

\bibitem{cs09}
A.~Corichi and P.~Singh,
\newblock {\em Geometric perspective on singularity resolution and uniqueness
  in loop quantum cosmology},
\newblock Phys. Rev. {\bf D80}, 044024 (2009).

\bibitem{Ashtekar:2018lag}
A.~Ashtekar, J.~Olmedo, and P.~Singh,
\newblock {\em Quantum transfiguration of Kruskal black holes},
\newblock Phys. Rev. Lett. {\bf 121}, 241301 (2018).

\bibitem{Ashtekar_2018}
A.~Ashtekar, J.~Olmedo, and P.~Singh,
\newblock {\em Quantum extension of the Kruskal spacetime},
\newblock Phys. Rev. {\bf D98}, 126003 (2018).

\bibitem{Giesel-Li-Singh2021}
K.~Giesel, B.-F. Li, and P.~Singh,
\newblock {\em Non-singular quantum gravitational dynamics of an LTB dust shell
  model: the role of quantization prescriptions},
\newblock arXiv:2107.05797.

\bibitem{abl}
A.~Ashtekar, M.~Bojowald, and J.~Lewandowski,
\newblock {\em Mathematical structure of loop quantum cosmology},
\newblock Adv. Theor. Math. Phys. {\bf 7}, 233 (2003).

\bibitem{Ashtekar_2005}
A.~Ashtekar and M.~Bojowald,
\newblock {\em Quantum geometry and the Schwarzschild singularity},
\newblock Class. Quant. Grav. {\bf 23}, 391 (2005).

\bibitem{Bohmer_2007}
C.~G. Bohmer and K.~Vandersloot,
\newblock {\em Loop quantum dynamics of the Schwarzschild interior},
\newblock Phys. Rev. {\bf D76}, 104030 (2007).

\bibitem{Campiglia_2008}
M.~Campiglia {\em et~al.},
\newblock {\em Loop quantization of spherically symmetric midi-superspaces: the
  interior problem},
\newblock AIP Conference Proceedings  (2008).

\bibitem{Gambini_2008}
R.~Gambini and J.~Pullin,
\newblock {\em Black holes in Loop Quantum Gravity: The complete space-time},
\newblock Phys. Rev. Lett. {\bf 101}, 161301 (2008).

\bibitem{Gambini_2013}
R.~Gambini and J.~Pullin,
\newblock {\em Loop quantization of the Schwarzschild black hole},
\newblock Phys. Rev. Lett. {\bf 110} (2013).

\bibitem{Gambini2014}
R.~Gambini, J.~Olmedo, and J.~Pullin,
\newblock {\em Quantum black holes in loop quantum gravity},
\newblock Class. Quant. Grav. {\bf 31}, 095009 (2014).

\bibitem{Corichi_2016}
A.~Corichi and P.~Singh,
\newblock {\em Loop quantization of the Schwarzschild interior revisited},
\newblock Class. Quant. Grav. {\bf 33}, 055006 (2016).

\bibitem{Olmedo:2017lvt}
J.~Olmedo, S.~Saini, and P.~Singh,
\newblock {\em From black holes to white holes: a quantum gravitational,
  symmetric bounce},
\newblock Class. Quant. Grav. {\bf 34}, 225011 (2017).

\bibitem{Gambini_2020}
R.~Gambini, J.~Olmedo, and J.~Pullin,
\newblock {\em Spherically symmetric loop quantum gravity: Analysis of improved
  dynamics},
\newblock Class. Quant. Grav. {\bf 37}, 205012 (2020).

\bibitem{Bojowald-singh2005}
M.~Bojowald, R.~Goswami, R.~Maartens, and P.~Singh,
\newblock {\em Black hole mass threshold from nonsingular quantum gravitational
  collapse},
\newblock Phys. Rev. Lett. {\bf 95}, 091302 (2005).

\bibitem{Goswami_2006}
R.~Goswami, P.~S. Joshi, and P.~Singh,
\newblock {\em Quantum evaporation of a naked singularity},
\newblock Phys. Rev. Lett. {\bf 96}, 031302 (2006).

\bibitem{Gambini_2009}
R.~Gambini, J.~Pullin, and S.~Rastgoo,
\newblock {\em Quantum scalar field in quantum gravity: the vacuum in the
  spherically symmetric case},
\newblock Class. Quant. Grav. {\bf 26}, 215011 (2009).

\bibitem{Bambi_2013}
C.~Bambi, D.~Malafarina, and L.~Modesto,
\newblock {\em Non-singular quantum-inspired gravitational collapse},
\newblock Phys. Rev. {\bf D88}, 044009 (2013).

\bibitem{Tavakoli_2014}
Y.~Tavakoli, J.~Marto, and A.~Dapor,
\newblock {\em Semiclassical dynamics of horizons in spherically symmetric
  collapse},
\newblock Int. J. Mod. Phys. D {\bf 23}, 1450061 (2014).

\bibitem{Ben_tez_2020}
F.~Benitez, R.~Gambini, L.~Lehner, S.~Liebling, and J.~Pullin,
\newblock {\em Critical collapse of a scalar field in semiclassical Loop
  Quantum Gravity},
\newblock Phys. Rev. Lett. {\bf 124}, 071301 (2020).

\bibitem{han2020improved}
M.~Han and H.~Liu,
\newblock {\em Improved effective dynamics of Loop-Quantum-Gravity black hole
  and Nariai limit}, 2020, arXiv:2012.05729.

\bibitem{Kelly_2}
J.~G. Kelly, R.~Santacruz, and E.~Wilson-Ewing,
\newblock {\em Black hole collapse and bounce in effective loop quantum
  gravity},
\newblock Class. Quant. Grav. {\bf 38}, 04LT01 (2020).

\bibitem{Bojowald:2008ja}
M.~Bojowald, T.~Harada, and R.~Tibrewala,
\newblock {\em Lema\^itre-Tolman-Bondi collapse from the perspective of loop
  quantum gravity},
\newblock Phys. Rev. {\bf D78}, 064057 (2008).

\bibitem{Israel1991}
C.~Barrab\`es and W.~Israel,
\newblock {\em Thin shells in general relativity and cosmology: The lightlike
  limit},
\newblock Phys. Rev. {\bf D43}, 1129 (1991).

\bibitem{goswami-joshi04}
R.~Goswami and P.~S. Joshi,
\newblock {\em Naked Singularity formation in scalar field collapse
  }, arXiv:gr-qc/0410144.

\bibitem{numlsu-1}
P.~Singh,
\newblock {\em Glimpses of spacetime beyond the singularities using
  supercomputers},
\newblock Comput. Sci. Eng. {\bf 20}, 26 (2018).

\bibitem{numlsu-2}
P.~Diener, B.~Gupt, and P.~Singh,
\newblock {\em Numerical simulations of a loop quantum cosmos: robustness of
  the quantum bounce and the validity of effective dynamics},
\newblock Class. Quant. Grav. {\bf 31}, 105015 (2014).

\bibitem{numlsu-3}
P.~Diener, B.~Gupt, M.~Megevand, and P.~Singh,
\newblock {\em Numerical evolution of squeezed and non-Gaussian states in loop
  quantum cosmology},
\newblock Class. Quant. Grav. {\bf 31}, 165006 (2014).

\bibitem{Meissner_2004}
K.~A. Meissner,
\newblock {\em Black-hole entropy in loop quantum gravity},
\newblock Class. Quant. Grav. {\bf 21}, 5245-5251 (2004).

\bibitem{Ashtekar_1998}
A.~Ashtekar, J.~Baez, A.~Corichi, and K.~Krasnov,
\newblock {\em Quantum Geometry and Black Hole Entropy},
\newblock Phys. Rev. Lett. {\bf 80}, 904-907 (1998).

\bibitem{Dreyer_2003}
O.~Dreyer,
\newblock {\em Quasinormal Modes, the Area Spectrum, and Black Hole Entropy},
\newblock Phys. Rev. Lett. {\bf 90}, 081301 (2003).

\bibitem{Domagala_2004}
M.~Domagala and J.~Lewandowski,
\newblock {\em Black-hole entropy from quantum geometry},
\newblock Class. Quant. Grav. {\bf 21}, 5233-5243 (2004).

\bibitem{Ansari_2008}
M.~H. Ansari,
\newblock {\em Generic degeneracy and entropy in loop quantum gravity},
\newblock Nucl. Phys. B {\bf 795}, 635-644 (2008).

\bibitem{Ashtekar_2009}
A.~Ashtekar,
\newblock {\em Loop quantum cosmology: an overview},
\newblock Gen. Rel. Grav. {\bf 41}, 707-741 (2009).

\bibitem{Ashtekar_2008}
A.~Ashtekar and E.~Wilson-Ewing,
\newblock {\em Covariant entropy bound and loop quantum cosmology},
\newblock Phys. Rev. {\bf D78}, 064047 (2008).

\bibitem{Bojowald2006}
M.~Bojowald,
\newblock {\em Loop quantum cosmology and inhomogeneities},
\newblock Gen. Rel. Grav. {\bf 38}, 1771-1795 (2006).

\bibitem{apsv} A.~Ashtekar, T.~Pawlowski, P.~Singh and K.~Vandersloot,
{\em Loop quantum cosmology of k=1 FRW models,}
Phys. Rev. D \textbf{75}, 024035 (2007)

\bibitem{ck1} A.~Corichi and A.~Karami, {\em Loop quantum cosmology of k=1 FRW: A tale of two bounces,}
Phys. Rev. D \textbf{84}, 044003 (2011)

\bibitem{ds1} J.~L.~Dupuy and P.~Singh, {\em Implications of quantum ambiguities in $k$=1 loop quantum cosmology: distinct quantum turnarounds and the super-Planckian regime,}
Phys. Rev. D \textbf{95}, no.2, 023510 (2017)

\bibitem{li-ps} B-F. Li, P. Singh, {\em Loop quantization prescriptions and black hole formation in the bound case }, {\em  In preparation} (2021). 

\end{thebibliography}

\end{document}